# Compact Pixelated Microstrip Forward Broadside Coupler Using Binary Particle Swarm Optimization

Kourosh Parsaei, *Student Member, IEEE*, Rasool Keshavarz, *Member, IEEE*, Rashid Mirzavand Boroujeni, *Senior Member, IEEE*, and Negin Shariati, *Senior Member, IEEE*

*Abstract*— In this paper, a compact microstrip forward broadside coupler (MFBC) with high coupling level is proposed in the frequency band of 3.5~3.8 GHz. The coupler is composed of two parallel pixelated transmission lines. To validate the design strategy, the proposed MFBC is fabricated and measured. The measured results demonstrate a forward coupler with 3 dB coupling, and a compact size of 0.12 $\lambda g \times 0.10 \lambda g$. Binary Particle Swarm Optimization (BPSO) design methodology and flexibility of pixelation enable us to optimize the proposed MFBC with desired coupling level and operating frequency within a fixed dimension. Also, low sensitivity to misalignment between two coupled TLs makes the proposed coupler a good candidate for near-field Wireless Power Transfer (WPT) application and sensors.

*Index Terms*— Binary particle swarm optimization, broadside forward coupling, flexible design method, pixelated near field coupler.

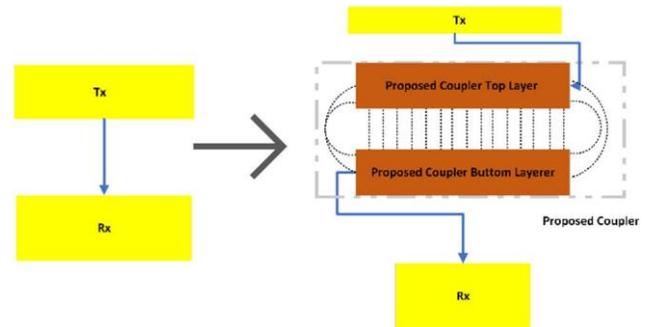

Fig. 1. Illustrates how the proposed coupler can be integrated into the systems such as [11] to provide near-field wireless communication or power transfer.

## I. INTRODUCTION

COUPLERS are a vital part of the majority of microwave and millimeter-wave systems where the signal or a portion of it is required to be transferred from one transmission line to another [1], [2]. In most cases, reducing the size and weight of the coupler is essential. Edge-coupling, as the most common technique in planar transmission lines, has been very well advanced and investigated [3], [4]. The main disadvantage of edge coupling is their limited coupling levels and sizeable footprint [3], [5]. The alternative technique is called broadside coupling which has recently drawn more attention [6]. The main advantages of broadside-couplers are the smaller size and higher coupling level between the resonators due to a larger effective area between transmitter and receiver [7]. Microstrip broadside couplers have been popular due to their relatively simple design method and fabrication and have been vastly used in many applications such as implant technologies, RFID, IoT devices, sensors, near-field wireless power transfer (WPT), etc. [8], [9], [10]. One of the features of the microstrip broadside couplers is the ability to have a wireless connection between the transmission and receiver sides. In many applications such as measurement devices, e.g. [11] the broadside microstrip couplers can be added as an integral part of the circuit to offer wireless power transfer for the measuring side. This feature adds flexibility to such measuring systems, sensors, or any other applications where a two-way wireless connection between one part and the rest of the circuit is advantageous. Fig.1 shows how the proposed structure can be used as the wireless power transfer for a system and replace the wired connection between the transmitter and the receiver part. This wireless connection can be used to transfer both data and power.

While the microstrip broadside couplers have been known for years, using the traditional analytical methods for designing a coupler that operates on a specific frequency range within a predefined size is challenging. In some applications the area where the coupler needs to be installed is fixed in size and, hence the classic approach where the coupler size changes with the operating frequency limits the degree of freedom in the coupler design [12], [13]. Therefore, investigating a flexible design methodology is tangible [14], [15].

Recently, new techniques based on AI, machine learning and optimization tools have been investigated in the design of RF and microwave circuits and systems to overcome the rigidness challenges classic methods face [16], [17].

An optimization tool is a mathematical way of solving an optimization problem [18], [19]. An optimization problem usually consists of the following parts:

1- Objective function (an equation to calculate the objective value).
2- A set of constraints (equalities or inequalities which all members of objective function must follow).
3- The variable boundaries (the limits on the range of variables) [20], [21].

An optimization tool can be continuous, integer or binary and the aim of the tool is to maximize/minimize the objective value considering the boundary and constraint rules [22], [23]. The most common optimization algorithms are Genetic Algorithm, Firefly Algorithm and Particle Swarm Optimization (PSO) [24], [25].

PSO has been one of the most used algorithms among electromagnetic researchers due to its effectiveness and ease of use [26]. By introducing transfer functions and a distinct location updating method, the binary version of particle swarm optimization (BPSO) was introduced by Kennedy and Eberhart [27]. The purpose of the transfer function is to plot the continuous realm parameter updating into binary mode. Since in binary mode the change in location does not have a physical meaning, the updating procedure purpose is to exchange between the two possible values for any given particle, zero (0) or one (1) [24].

In the proposed structure, the BPSO was used combined with the pixelation technique. The pixelation means to assume breaking down an area into small size pieces (in the proposed structure, rectangle shapes) [28]. The point of pixelating an area is to define those pixels or small areas in the structure as particles in the optimization algorithm which enable us to look at that area in a binary manner. The outcome of the algorithm is a binary string of zeros and ones which determines the formation of pixelated areas that satisfy the goal defined for the BPSO [24].

In this paper, a highly efficient compact Microstrip Forward Broadside Coupler (MFBC) is designed, implemented and optimized based on pixelation and BPSO technique which can be used in near-field WPT systems, implantable biodevices, measuring systems etc. Fig. 2. illustrates the design method flexibility and the parameters which the optimization tool can optimize. The major contributions of this paper are summarized as follows:

- The proposed methodology enables us to optimize the MFBC's desired parameters (in this structure it's the coupling level as the goal of PSO while maintaining the bandwidth).
- This technique is very flexible and can be applied on most RF and microwave structures (filters, duplexers, antennas, etc.) at different frequencies and sizes.
- Low sensitivity to misalignment between two coupled TLs is achieved. Hence, this coupler is a good candidate for various practical measurement systems for WPT scenarios, IoT devices, and implantable biodevices.

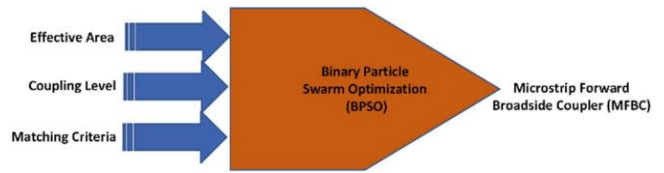

Fig. 2. Proposed design methodology.

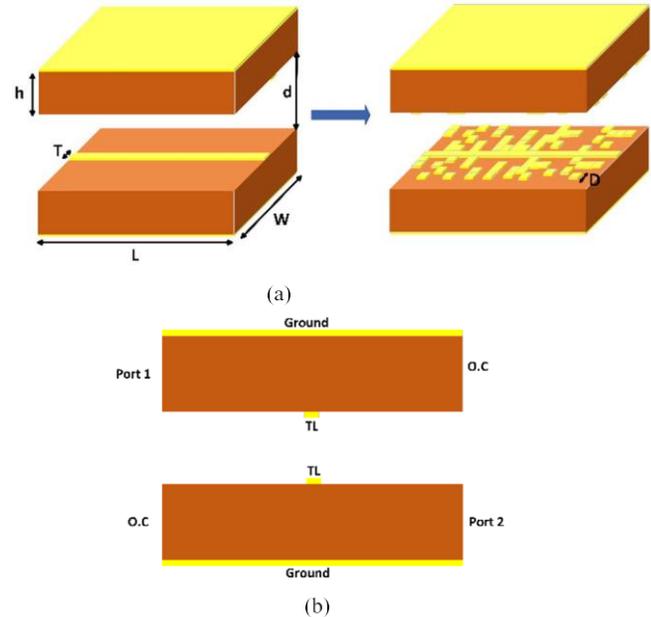

Fig. 3. The initial and pixelated coupler. a) 3D view of the initial coupler with no pixels on the left, and 3D of the pixelated structure on the right side. (b) 2D view of the initial coupler without pixels.

- Ease of integration to other parts of the circuit due to developed low profile and microstrip type structures.

The organization of this paper is as follows: Coupler design and methodology are presented in Section II. The proposed MFBC performance is validated by analytical, simulation, and measurement results in Section III. Lastly, the conclusion is presented in Section IV.

## II. COUPLER DESIGN METHODOLOGY

According to [7], a parallel four-port coupled lines with two excitation ports and the other two ports open ended present a bandpass frequency response. The initial coupler which is shown in the left side of Fig. 3 (a) is designed based on this concept. The proposed structure includes two microstrip transmission lines (TLs) which are positioned parallel to each other. There are four ports, two are open circuits (O.C.) ends, and the other two ports are designed for the input/output signals as the excitation ports (Fig. 3(b)). The area on top and below both transmission lines are appointed for the pixelation technique where the optimization is applied.

In Section A, the conventional coupler design method and its challenges are explained. In Section B, the basis of the proposed method, the flexibility, and how it can help address the challenges in the conventional method are explained.

### A. Forward Broadside Coupler Design

The coupled microstrip lines shown in Fig. 3 (a) are symmetric, where the two conducting strips have the same width and position relative to the ground planes. Therefore, the electrical characteristics of the coupled lines can be determined based on the effective capacitances between the lines and the velocity of propagation on the line.

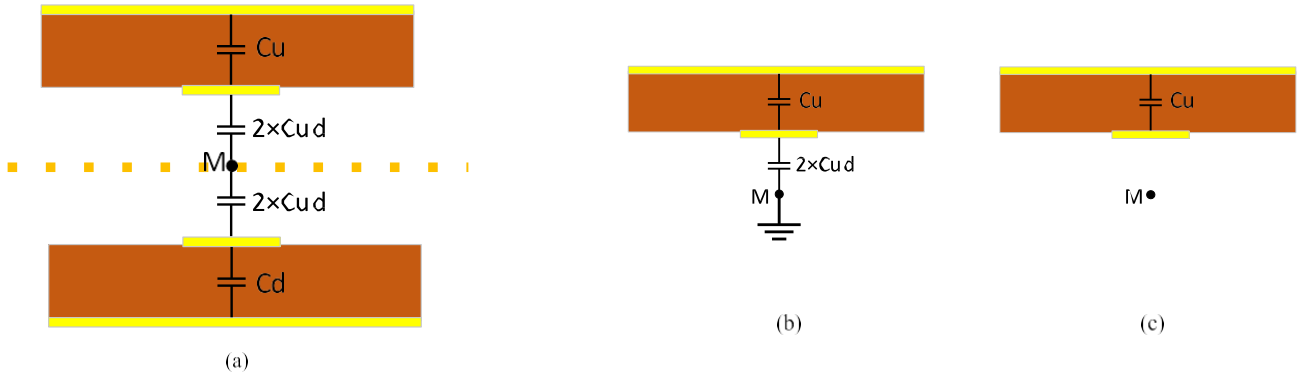

Fig. 4. a) Capacitance model of two coupled TLs, and capacitance representation for single TL in, b) odd mode, c) even mode.

As depicted in Fig. 4, $C_{ud}$ represents the capacitance between the strip conductors of up and down TLs and $C_u$ and $C_d$ represent the capacitances between up and down strip conductors to the related grounds, respectively. Since the up and down strip conductors are identical in size and location relative to the ground conductor, it can be concluded that $C_u = C_d$.

The interaction between the two transmission lines (TLs) in a microwave coupler results in two different modes of wave propagation: the even mode and the odd mode. The even mode is characterized by the fact that the current in both transmission lines is in the same direction.

On the other hand, the odd mode occurs when the current in the two transmission lines is in opposite directions. It is important to note that the even and odd modes are orthogonal to each other, meaning that there is no coupling between them. This orthogonality allows the two modes to be independently analysed and optimized, leading to improved performance of the microwave coupler [7].

In the equivalent circuit model of conventional TLs, the coupling between them is represented by mutual inductances and mutual capacitances, which can be adjusted to control the coupling level between the two lines. In microstrip transmission lines, mutual inductance is often negligible compared to the capacitive coupling between the lines. This is because microstrip lines are typically constructed on a dielectric substrate with a relatively large height above a ground plane. The relatively large height above the ground plane reduces the mutual inductance between the lines and enhances the capacitive coupling between them.

For the even mode, the electric field has even symmetry around the dashed line in Fig. 4 (c), and no current flows between the two strip conductors. This leads to an open circuit at point M and $C_{ud}$ is effectively open-circuited. The resulting capacitance of either line to ground for the even mode is:

$$C_e = C_d = C_u = C \quad (1)$$

Therefore, the characteristic impedance for the even mode is:

$$Z_{0e} = \sqrt{\frac{L}{C_e}} \quad (2)$$

where $L$ is the inductance of each TL (up or down).

For the odd mode, the electric field lines have an odd symmetry around the dashed line, and a voltage null or ground exists at point M. In this case the effective capacitance between both strip conductors and their ground is:

$$C_o = C_d + 2C_{ud} = C_u + 2C_{ud} \quad (3)$$

So, the characteristic impedance for the odd mode is:

$$Z_{0o} = \sqrt{\frac{L}{C_o}} \quad (4)$$

An arbitrary excitation of a coupled line can always be treated as a superposition of appropriate amplitudes of even- and odd-mode excitations.

If we define the coupling coefficient, $C_c$ as [7]

$$C_c = |S_{21}| = \sin(\theta) = \sin\left(\frac{\beta_e - \beta_o}{2}l\right) \quad (5)$$

where

$$\beta_e = \omega\sqrt{LC_e} = \omega\sqrt{LC} \quad (6)$$

$$\beta_o = \omega\sqrt{LC_o} = \omega\sqrt{L(C + 2C_{ud})} \quad (7)$$

and

$$\beta_o - \beta_e = \omega\sqrt{L}\left(\sqrt{(C + 2C_{ud})} - \sqrt{C}\right) \quad (8)$$

The maximum coupling level occurs at $\theta = \pi/2$.

$$\frac{\beta_e - \beta_o}{2}l = \theta = \frac{\pi}{2} \quad (9)$$

Consequently, the coupler length to achieve maximum coupling level is:

$$l_{max} = \frac{\pi}{\beta_e - \beta_o} \quad (10)$$

According to (10), by increasing the difference between even and odd modes propagation constants, $\beta_e$ and $\beta_o$, the $l_{max}$ will be decreased. Fig. 5 shows the effect of effective coupling area on $l_{max}$. As shown in this figure, by increasing the initial effective coupling area ($A_e$) to $A'_e$, the new coupler length ($L'_{max}$) is decreased compared to the initial value ($L_{max}$). Hence, by increasing the effective coupling area in the broadside coupler the coupler length is decreased which leads to achieving a compact coupler without losing the coupling level. However, while increasing the effective area leads to a more compact coupler, the TL at each side has to be matched to the port impedance (50 $\Omega$). It is shown in Section B that using the pixelated technique can satisfy both matching conditions and increase the effective coupling.

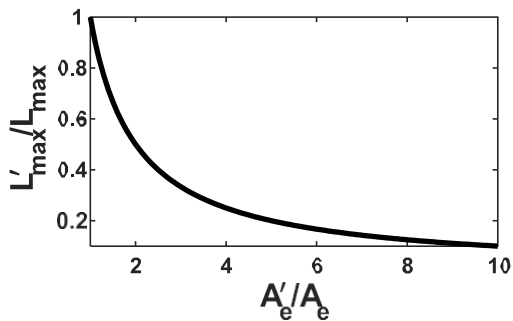

Fig. 5. Effective area influence on the maximum coupling length.

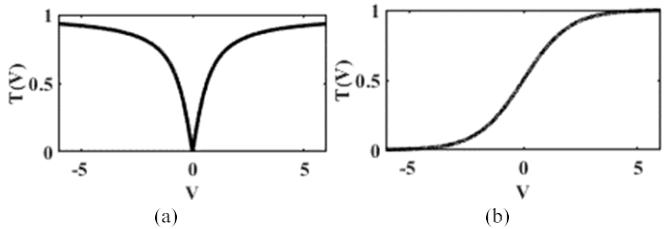

Fig. 6. Transfer Function plots (a) V-shaped (b) S-shaped.

### B. Flexible Design Algorithm BPSO

In PSO, as an optimization tool for real-number space, the assumed solution to the subject is indicated as a particle on the location Xi (i is the iteration number) and velocity of Vd (d is the direction) in a multidimensional space. Each particle keeps records of its own Personal Best (Pbest) performance. In the next iteration (i = i + 1), the algorithm calculates the next position based on its current direction, its own Pbest, and the best performance observed by any particle which is called Global Best (Gbest). This process continues until the optimization pre-defined stopping rules are met.

By introducing transfer functions and a distinct location updating method, the binary version of particle swarm optimization was introduced [27]. The transfer function determines the probability of each particle's value to be 1 or 0 based on the previous iteration's results. There are two main transfer function families, S-shaped and V-shaped which are demonstrated in Fig. 6.

In this paper, one of each transfer function families are used to compare their efficiency. The main difference between the two transfer function families is that in V-shaped function the algorithm does not determine a definitive 0 or 1 value for the particle and allows the particle to stay at its current status or take a complementary value. Whereas the S-shaped transfer function determines if the particle value is a zero or one at each iteration. In [22], both transfer functions are presented in detail including their advantages, disadvantages, and applications.

For each particle at location Xi, where the Gbest and Pbest are known, the velocity at the next iteration ($v^{t+1}_i$) is:

$$v^{t+1}_i = wv^t_i + c_1 \times rand \times \left(pbest_i - x^t_i\right) + c_2 \times rand \times (gbest - x^t_i) \quad (11)$$

It is worth mentioning, $w$ is the weighting function, $c_1$ and $c_2$ are acceleration coefficients. Depending on which transfer function is being used, the algorithm calculates the new formation of the pixelated area.

The pixelated area is a continuous area in the structure where the optimization takes place and changes in that area have an impact on the optimization objective parameter. In an ideal situation, it is desirable that changes in the pixelated area do not affect the other parameters of the structure. Once the area is specified, it is presumed to break down into small pixels or cells. In our proposed MFBC, there are four pixelated areas, the space above and under the transmission line (TLs) in both parts of the coupler. Each space is broken into 30 columns and 14 rows, in a total of 420 pixels or particles. Any of the pixels can have a value of zero or one assigned to them. A value of 1 means that the small area allocated to that pixel is covered with metal, and consequently zero means a non-metal cell. By changing the values assigned to the pixels on the transmission and receiver side, the response of the coupler changes.

Fig. 7 illustrates the design procedure of the proposed MFBC. The defined goal for the optimization tool is to reduce the operating frequency to half while maximizing the coupling level ($S_{21}$). The initial values for the pixels are randomly generated. Personal Best (Pbest) and Global Best (Gbest) are initially set to be minus infinity.

Fig. 8 illustrates the step-by-step procedure of the proposed method. At each iteration, the optimization algorithm in MATLAB updates the values for the pixelated area and the Computer Simulation Technology (CST) run the simulation on the structure with new pixelation formation and returns the S parameters back to the MATLAB. The algorithm extracts the $S_{21}$ at the desired frequency and based on BPSO formulas, updates the Pbest, Gbest, pixels values and requests CST to simulate the structure with the new values. Every time the algorithm changes the values of the particles, the structure in the pixelated area changes accordingly. The rest of the structure remains unchanged.

A fundamental advantage of the proposed method is the ease of applying pixelation and BPSO in different systems and locations within the structure. It enables the customization to be based on the application's needs without going through onerous calculations. For instance, for a coupler, coupling level, operating frequency and return loss can be enhanced or the dimension can be reduced (please see Fig. 9). It is worth mentioning, the flexibility of this method is only available during the design process and cannot be tuned after fabrication.

### III. SIMULATION, MEASUREMENT AND DISCUSSION

As mentioned in Section II-B, the BPSO algorithm is coded in MATLAB software to design the proposed MFBC, and CST electromagnetic simulator software is used for the full wave simulation to obtain the S parameters. The goal for the BPSO algorithm is to optimize the coupling level at the center frequency. To minimize the risk of the algorithm sticking to a local maximum, the maximum iteration number is set high (100 iterations) to give the tool the chance to explore different compositions. Also, a dynamic weighting function (0.9 at the first iteration and gradually reduced by each iteration to 0.4) is used to reduce the conversion rate at lower iteration numbers. This allows the algorithm to explore more formations. The acceleration coefficients $c_1$ and $c_2$ are both set to 2.

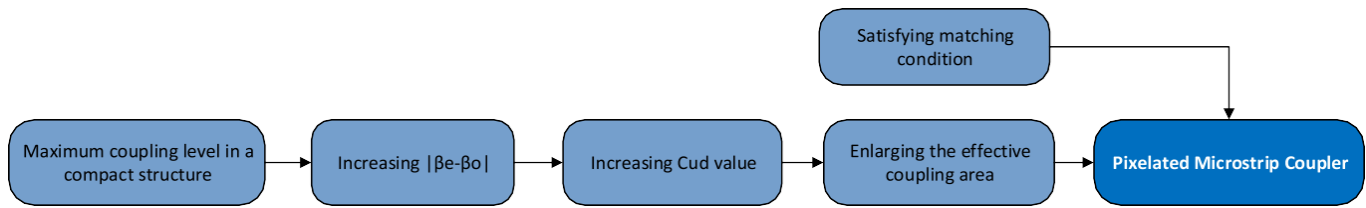

Fig. 7. Design procedure of the proposed MFBC using pixelation and BPSO optimization tool.

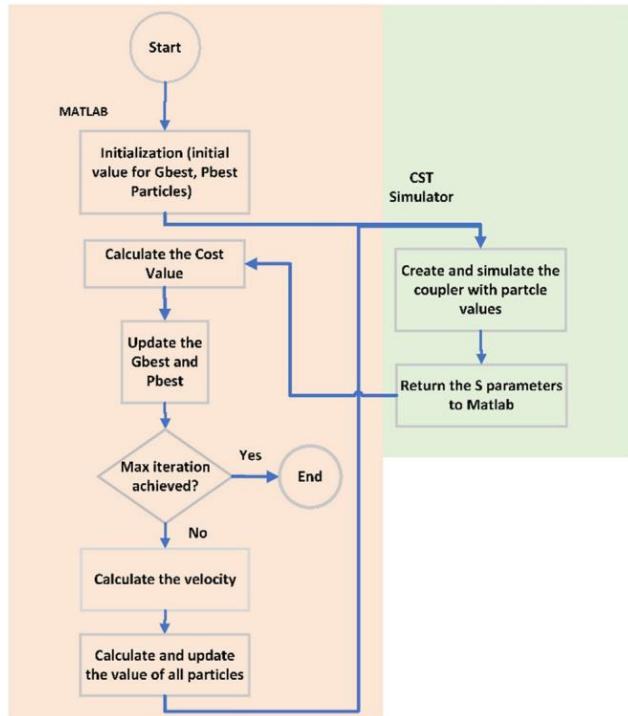

Fig. 8. Design flowchart of the proposed coupler.

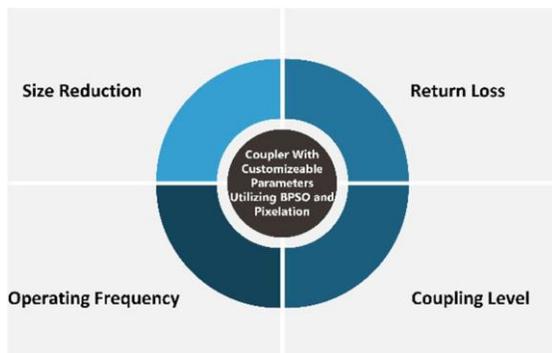

Fig. 9. Benefits of using pixelation technique with PSO for designing coupler.

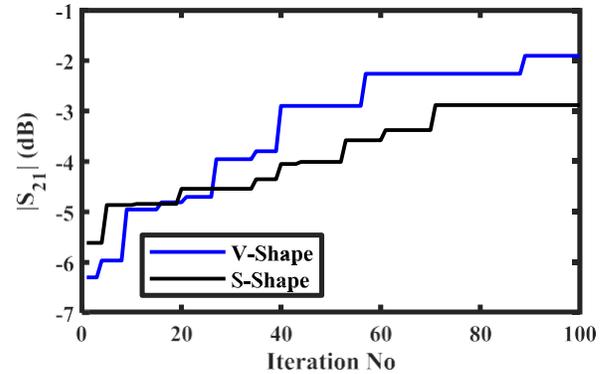

Fig. 10. Comparison of convergence curves obtained by V-shaped and S-shaped transfer functions.

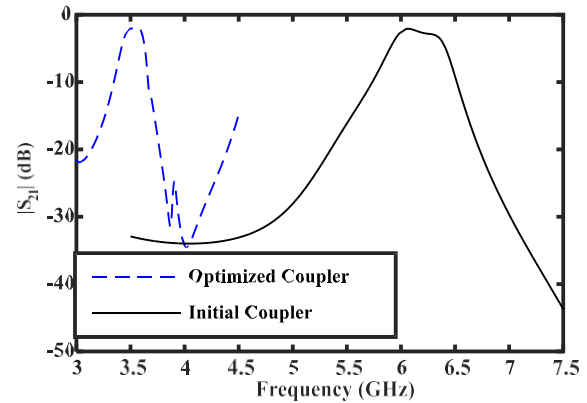

Fig. 11. A comparison between the simulation results of the initial structure with no pixels and the optimized pixelated structure.

Fig. 10 shows the convergence pattern of V-shape versus S-shape functions. The defined fitness value indicates the coupling level at the center frequency of 3.5 GHz. There are two conditions defined for the algorithm to stop the process, either achieve a coupling level higher than −2 dB or finish the 100 iterations and save the best results as the optimized structure. As it can be seen in Fig. 10, with the S-shape function a coupling level higher than −2 dB is never achieved so the algorithm continued until the $100^{th}$ iteration. However, with the V-shaped function on the 90th iteration, the coupling level higher than −2 dB is achieved so the algorithm stopped and provided the final structure as the optimized version. It should be noted that for different applications and designs, transfer functions may perform differently. In the proposed structure, V-shape function outperforms the S-shape, but this cannot be generalized. For analysis and measurement purposes, the structure achieved by the V-shape algorithm is fabricated and presented in this paper.

Fig. 11 compares the simulation coupling level of the initial structure with no pixels and pixelated structure. As it can be seen, the coupling level is slightly improved and the operating frequency is shifted from 6 to 3.5 GHz on the same structure size and shape with the addition of pixels.

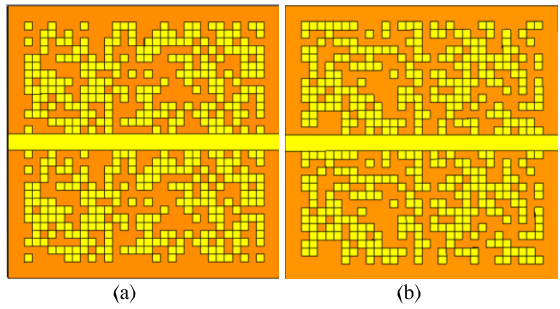

Fig. 12. The pixelated pattern generated by the BPSO algorithm. (a) Top view 3.5 GHz (b) Top View 3.8 GHz.

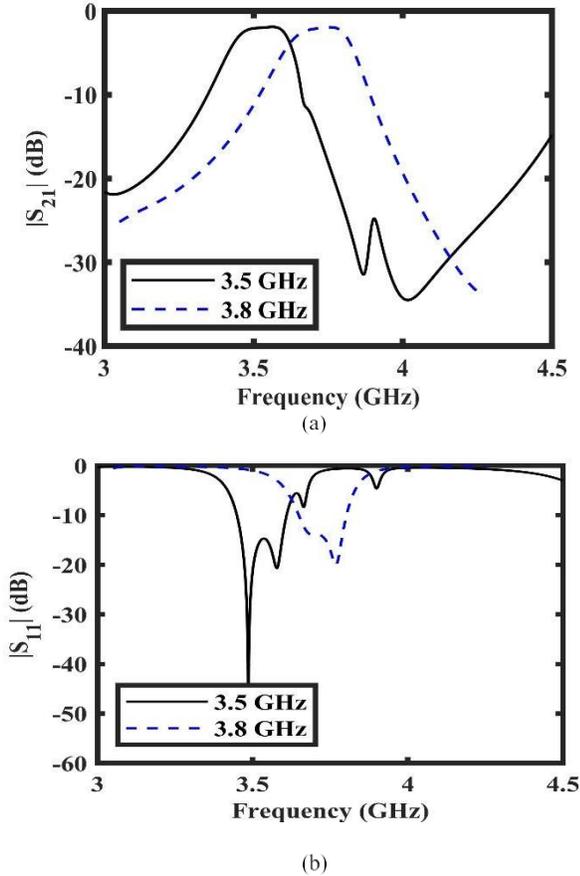

Fig. 13. Simulation results (a) $S_{21}$ of the structure in 3.5 and 3.8 GHz (b) $S_{11}$ in 3.5 and 3.8GHz.

The operating frequency is the frequency that BPSO in MATLAB is coded to optimize the coupling level. By changing the operating frequency defined for the BPSO in MATLAB, the tool will change the composition of the pixels on the top layer until it finds the optimized structure that operates in the new frequency. In order to demonstrate this, the structure is optimized in two different frequencies, 3.5 and 3.8 GHz. Fig. 12 (a) and 12 (b) show the formation of the optimized patch view in 3.5 and 3.8 GHz respectively. Fig. 13(a) and 13(b) show the coupling level and return loss at each frequency. A key point to consider using this technique is that every time the optimization starts, the BPSO starts with a random pixelation formation as the initial state and applies the optimization until the goals defined for the optimization tool are met. This means that if we rerun the simulation with the same optimization goals to the exact same structure, the final structure of the pixels can be varied, and the results can be slightly different. However, the results follow the specified goals within a defined threshold.

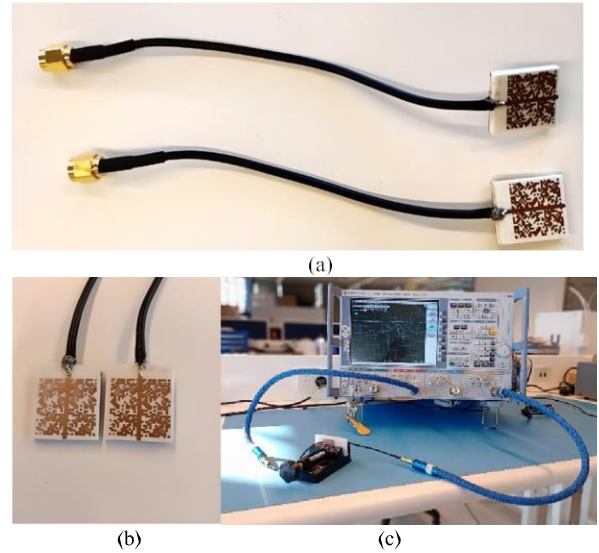

Fig. 14. The fabricated setup (a) Coaxial feed connection (b) Top view of the fabricated pixelated structure on RO-4003 (c) measurement setup.

TABLE I
THE DIMENSION OF THE MFBC STRUCTURE ACCORDING TO THE PARAMETERS IN FIG.3

| Parameter | W | L | T | D | h | d |
|---|---|---|---|---|---|---|
| Value (mm) | 20 | 17 | 1 | 0.5 | 3.2 | 5 |

The proposed MFBC is fabricated on Rogers4003C substrate, with $\varepsilon_r$ = 3.55 and loss tangent of 0.0025 (Fig. 14(b)). The dimension of the structure is given in Table I. Due to small size of the pixels and sharp edges; it is important to use a precise machine for fabrication.

Because of the small size and space between the two sides of the proposed coupler, it was impractical to use any commercial SMA connector, not even the miniaturized ones. The body of the SMA connector introduces a parasitic element that disrupts the S parameters, thereby impacting the overall performance. Instead of using the connector, coaxial cables are connected to port 1 and port 2 as input and output ports, and the other ports are open circuits (Fig. 14 (a)).

The measurement setup is shown in Fig. 14 (c). The S-parameter curves are measured using Rohde and Schwarz ZVA40 Vector Network Analyzer. The simulated vs measured results at 3.5 GHz center frequency are shown in Fig. 15. As it can be seen, the coupler operates at the frequency range of 3 to 3.6 GHz (600 MHz bandwidth) with a maximum $S_{21}$ of -4.13 dB. The measurement results are aligned with simulations.

Another feature of the proposed structure is that it exhibits low sensitivity to misalignment and rotation of one side with respect to the other TL. As it was discussed in Section II-A pixelation changes the effective area on the patch. Rotating the structure with pixels should not change the response significantly. Fig. 16 (b) and (c), show the simulation and measured results of rotating one side of the coupler by $\alpha$= 0˚, 45˚, 90˚, where $\alpha$ is the misalignment angle (Fig. 16(a)). The measurement results however show a reduction in the bandwidth with rotation angle increases. The slight difference between the measurement and simulation outcomes could be attributed to the challenges in preserving the proper angle and distance during measurement. Additionally, the measurement results could exhibit reduced bandwidth due to the close proximity of the excitation ports. When the rotation angle is increased, the excitation ports come closer, causing a deviation in the bandwidth. Considering the small size and distance between the two sides, the measurement results are within an acceptable range. The low sensitivity to misalignment makes the proposed MFBC advantageous for applications where the rotation of the TLs is inevitable.

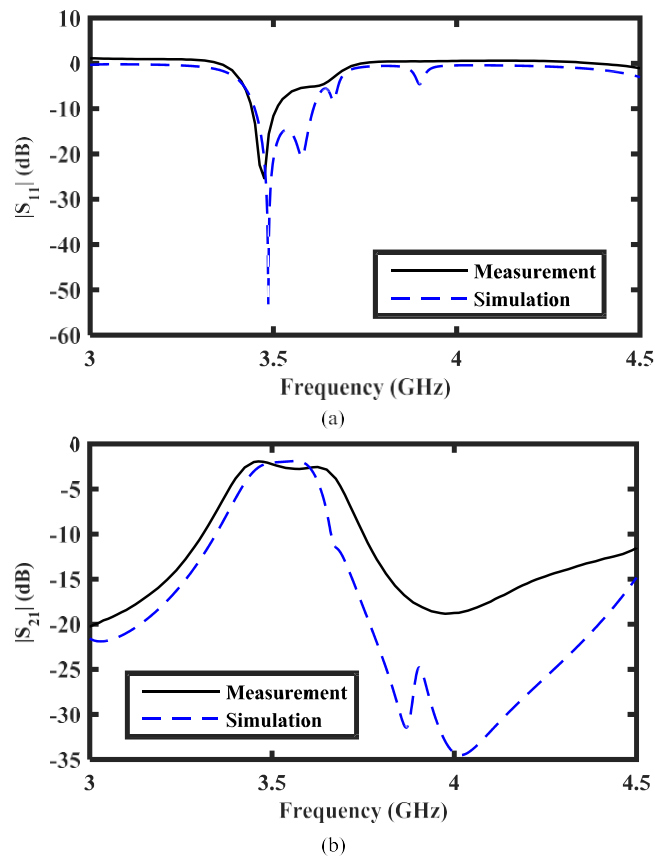

Fig. 15. S-parameters of the fabricated structure versus the simulation results. (a) $S_{11}$ (b) $S_{21}$.

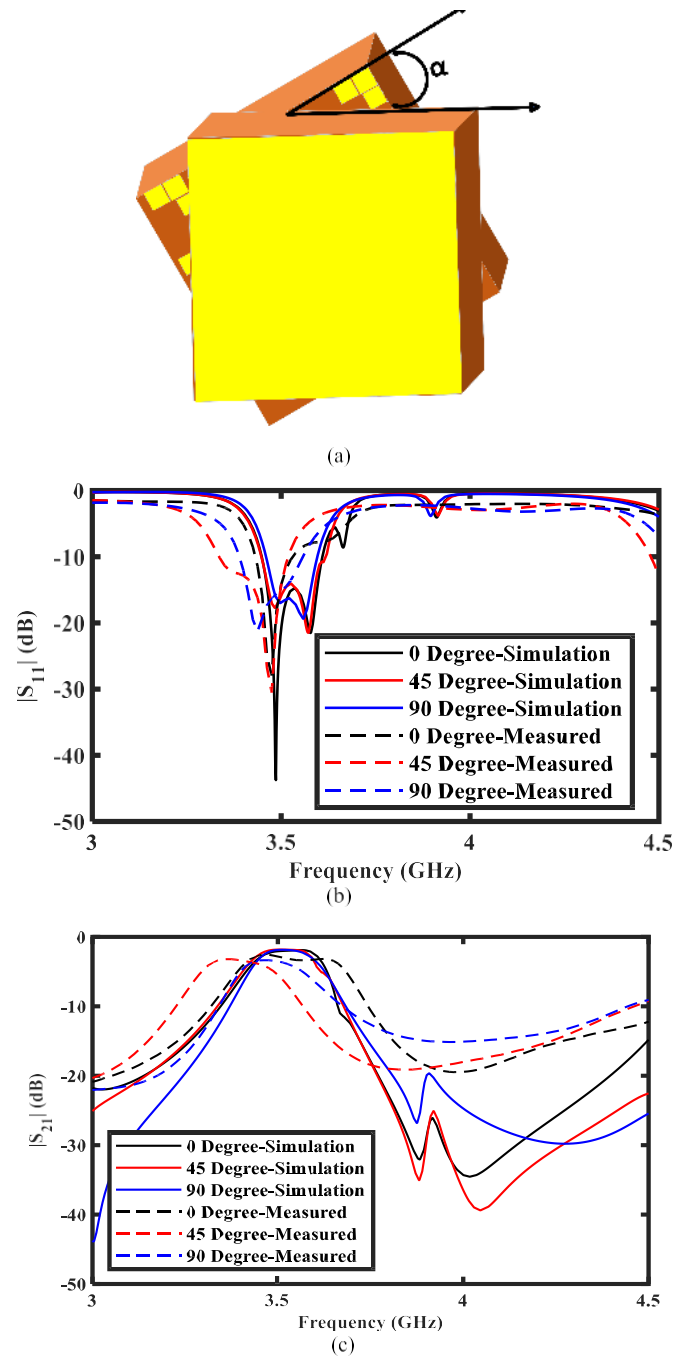

Fig. 16. Rotating the transmitter part of the coupler (a) 3D view with $\alpha$ degree rotation (b) Simulation and measurement S11 results (c) Simulation and measurement S21 results.

Fig. 17 illustrates the effect of changing the distance between the two TLs. The proposed MFBC is optimized to operate when the distance is 5 mm. While for some measurements (for instance, Proximity sensors, Capacitive or Magnetic sensing) or WPT applications, this is a valid distance, 5mm as the initial distance is arbitrary. Please note that the algorithm is flexible and can be adapted to different applications, hence, it can perform the optimization at any other defined distance during the design process.

As it can be seen in Fig. 17(a) and (b), changing the distance in any direction causes a gradual deterioration in the performance of the proposed MFBC. Reducing the distance from 5 mm to 4 and 3 mm causes distortion in impedance matching. Increasing the distance, however, cause a reduction in the coupling level and bandwidth.

Fig. 18(a) shows the measured S21 results of the proposed MFBC whilst changing the distance between the sides. Fig. 18(b) shows the coupling level comparison between simulated and measured results at the frequency of 3.5 GHz with changing distance between the two sides. The simulated and measured results show that increasing distance between the two sides results in a reduction in the coupling level. The slight difference between the measured and simulated results is due to human error and a lack of precise measurement setup. Table II compares the proposed MFBC with other couplers in recent literatures. For couplers operating around 3.5 GHz, the size of the proposed MFBC is considerably smaller which is in line with the change in effective area in the proposed structure using the pixelation method. As it can be seen in Table II, except [29], the proposed MFBC's coupling level is higher than others. Considering [29] is larger in size, operates at a higher frequency and provides lower bandwidth, the proposed coupler shows better performance.

TABLE II
COMPARISON TABLE

| Ref. | Technique | Freq. (GHz) | Bandwidth (%) | Coupling Level (dB) | Dimension ($\lambda_g \times \lambda_g$) | Rotation Capability |
|---|---|---|---|---|---|---|
| [29] | circular sector patch | 30 | 11.2 | 1.04 | 1.3 × 1.1 | NA |
| [30] | Slotted-Microstrip Line | 2.4 | 52 | 2.8 | Not Given | NA |
| [31] | Fragment-Type Structure | 2 | 45 | 20 | 0.25× 0.1 | NA |
| [32] | open stubs | 1.265 | 54 | NA | 0.53× 0.34 | NA |
| [32] | shorted stubs | 1.35 | 66 | NA | 0.6 × 0.27 | NA |
| [5] | SPC 6-line | 2.2 | 56.67 | 3.05 | 0.034 | NA |
| [33] | phase reconfigurable synthesized transmission lines (PRSTLs) | 2.4 | 10.3 | 1.2 | 0.65× 0.27 | NA |
| This work | Pixelation & BPSO | 3.5 | 18 | 1.13 | 0.12 × 0.10 | Yes |

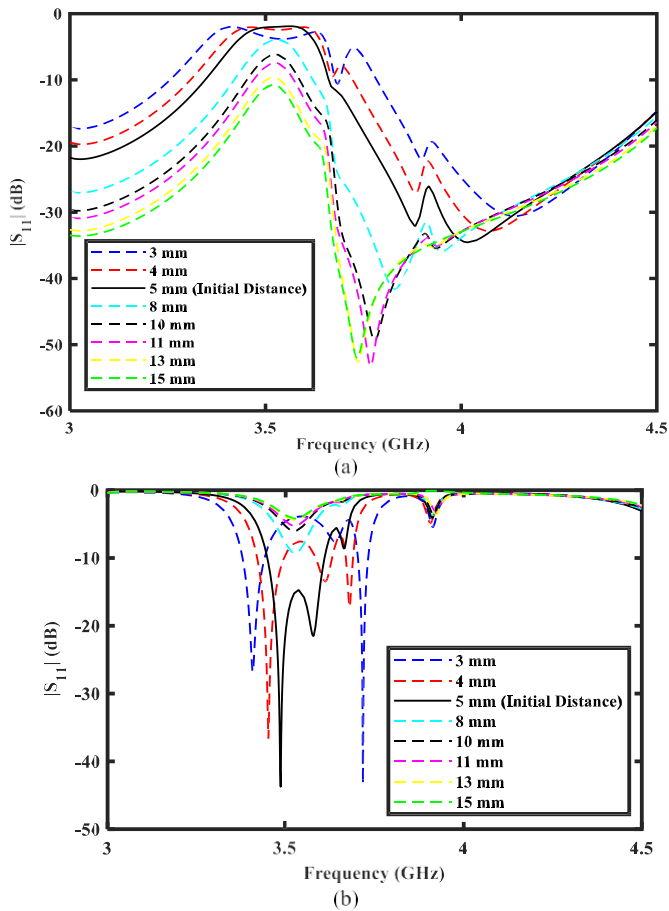

Fig. 17. Simulated S parameters with sweeping the distance between the TLs (a) S21 (b) S11.

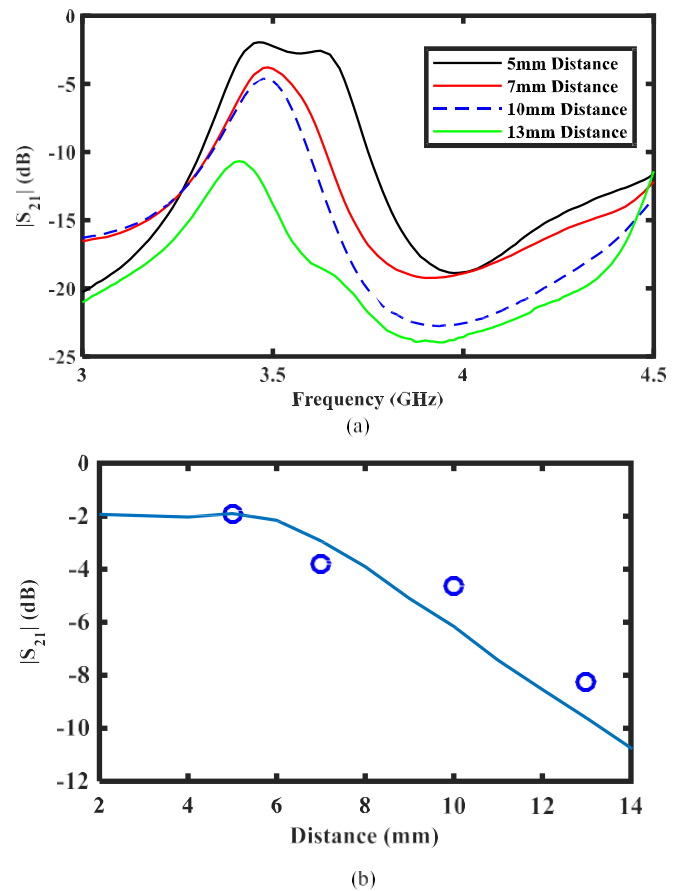

Fig. 18. Impact of increasing distance on S21 (a) Measured results (b) Comparison between measured and simulated results, line represent the simulation results, the dots shows the measured results.

## IV. CONCLUSION

A compact microstrip broadside forward coupler with high coupling level based on pixelation and BPSO algorithm is presented. The simulation and measurement results show an increase in the coupling level, a reduction in the resonant frequency and effectively reduction in the size of the coupler structure without sacrificing any other attribute of the conventional microstrip coupler. In order to demonstrate the flexibility of the proposed method, the coupler was optimized at two different frequencies by only changing the formation of the pixels.

The measured results of the fabricated prototype show a frequency range from 3.2 to 3.8 GHz and high coupling coefficient of 3 dB at the center frequency. The flexibility of pixelation and BPSO design methodology enable us to design and optimize microwave devices with desired performance within a predefined dimension.

## ACKNOWLEDGMENT

Kourosh Parsaei would like to thank the receipt of a Top-Up Postgraduate Scholarship from Food Agility CRC. The CRC program supports industry-led collaboration between industry, research, and the community.

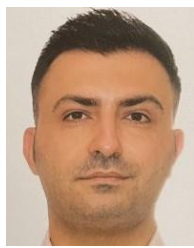

**Kourosh Parsaei** (Student Member, IEEE) received the Associate Diploma degree in electrical engineering from Shahid Bahonar University, Kerman, Iran, and the bachelor's degree in telecommunication engineering from APU University, Malaysia. His primary research interests include wireless power transfer (WPT), near field communication (NFC), sensor technologies, and optimization tools.

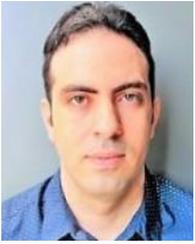

**Rasool Keshavarz** (Member, IEEE) was born in Shiraz, Iran, in 1986. He received the Ph.D. degree in telecommunications engineering from the Amirkabir University of Technology, Tehran, Iran, in 2017. He is currently a Senior Research Fellow with the RFCT Laboratory, University of Technology, Sydney, Australia. His main research interests are RF and microwave circuit and system design, sensors, antenna design, wireless power transfer (WPT), and RF energy harvesting (EH).

He received various awards, such as the Best AUT M.Sc. Researcher in 2007, the Best AUT Ph.D. Researcher in 2011, the Best MICT National Researcher in 2013, the National Elite Foundation Young Professor Grant in 2014, the AITF Elite PDF in 2015, the Honorable CMC Industrial Collaboration in 2017, the TEC Edmonton Innovation in 2019, the CMC Industrial Collaboration in 2021 as a Supervisor, the UofA Innovation in 2021, and three UofA Innovation in 2022. He is a registered member of the Association of Professional Engineers and Geoscientists of Alberta.

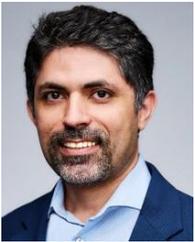

**Rashid Mirzavand Boroujeni** (Senior Member, IEEE) received the B.Sc. degree from the Isfahan University of Technology, Isfahan, Iran, in 2004, and the M.Sc. and Ph.D. degrees from the Amirkabir University of Technology (Tehran Polytechnic), Tehran, Iran, in 2007 and 2011, respectively, all in electrical engineering. He is currently an Assistant Professor with the Department of Electrical and Computer Engineering, University of Alberta, Edmonton, AB, Canada, where he leads the Intelligent Wireless Technology Group. He is also an Adjunct Fellow with the Faculty of Engineering and IT, University of Technology Sydney, Australia. He serves as the Specialty Chief Editor for *Frontiers in the Internet of Things* (IoT Enabling Technologies Section). He has three granted and eight filled U.S. patents and is the (co)author of more than 100 journals and 70 conference papers. His major research interests include, but are not limited to, RF/microwave/mm-wave circuits, sensors, reconfigurable intelligent surfaces and antennas, numerical methods, and measurement systems.

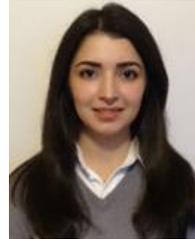

**Negin Shariati** (Senior Member, IEEE) received the Ph.D. degree in electrical-electronics and communication technologies from the Royal Melbourne Institute of Technology (RMIT), Australia, in 2016.

She worked in industry as an electrical-electronics engineer from 2009 to 2012. She is currently a Senior Lecturer with the School of Electrical and Data Engineering, Faculty of Engineering and IT, University of Technology Sydney (UTS), Australia. She established the State-of-the-Art RF and Communication Technologies (RFCT) Research Laboratory, UTS, in 2018, where she is currently the Co-Director and leads research and development in RF technologies, sustainable sensing, energy harvesting, low-power Internet of Things, and AgTech. She leads the Sensing Innovations Constellation at Food Agility Corporative Research Centre (CRC), enabling new innovations in agriculture technologies by focusing on three key interrelated streams; sensing, energy, and connectivity. She is also the Director of Women in Engineering at IT (WiEIT), Faculty of Engineering and IT, driving positive change in gender equity and diversity in engineering and IT. Since 2018, she has been holding a joint appointment as a Senior Lecturer with Hokkaido University, externally engaging with research and teaching activities in Japan. Her research interests are in RF-electronics circuits and systems, sensors, antennas, RF energy harvesting, simultaneous wireless information and power transfer, and wireless sensor networks.

Dr. Shariati was a recipient (2021) and a finalist (2022) of standout research awards in the IoT Awards Australia and IEEE Victorian Section Best Research Paper Award in 2015. She attracted more than six million dollars worth of research funding across a number of ARC, CRC, industry, and government-funded research projects over the past three years, where she has taken the lead a chief investigator (CI) role and also contributed as a member of the CI team.